\documentclass[]{aa}
\usepackage{graphics}
\usepackage{psfig}
\begin{document}
\def\teff{$T\rm_{eff }$}
\def\lambo{$\lambda$ Boo }

\title{
A speckle 
interferometry survey of $\lambda$ Bootis stars 
\thanks{Based on
observations made with the 
Italian Telescopio Nazionale Galileo (TNG) operated on the
island of La Palma 
by the Centro Galileo Galilei of the CNAA (Consorzio Nazionale per
l'Astronomia e l'Astrofisica) 
at the Spanish Observatorio del Roque de los Muchachos of
                            the Instituto de Astrofisica de Canarias }}

   \subtitle{}

\author{ Enrico \,Marchetti \inst{1,2}
\and Rosanna \,Faraggiana \inst{3}
\and Piercarlo \,Bonifacio \inst{4}
          }

  \offprints{E. Marchetti}

\institute{
European Southern Observatory, Karl--Schwarzschild--Str. 2, D-85748 
Garching bei M\"unchen, Germany
\and
Centro Galileo Galilei, 
Apartado de Correos 565, E-38700 Santa Cruz de la Palma,
Canary Islands, Spain
\and 
Dipartimento di Astronomia, Universit\`a degli Studi di Trieste,
Via G.B.Tiepolo 11, I-34131 Trieste, Italy 
\and Osservatorio Astronomico di Trieste, Via G.B.Tiepolo 11, 
I-34131 Trieste, Italy 
}

\authorrunning{Marchetti et al.}
\mail{emarchet@eso.org}

\titlerunning{TNG speckle observations of \lambo stars}

\date{Received 18 December 2000 / Accepted 12 February 2001}

\abstract{
A search for duplicity of \lambo stars has been made by
using the speckle camera installed at the Telescopio Nazionale
Galileo.
The operation mode and the reduction procedure allow one
to obtain not only the separation, but also the magnitude difference
between the components; the latter parameter is fundamental
for determining the degree of contamination from the secondary
component of a binary system and thus the importance of the
veiling effect that produces absorption lines weaker than normal.
Two stars, HD 38545 and HD 290492, are close binaries with
values of the separation and of the magnitude difference such
that  only a composite spectrum can be observed.
For another 15 \lambo candidates, observed with negative results,
the upper limits of a possible companion separation are given.
\keywords{               
               Techniques: interferometers -
               Stars: chemically peculiar -
               binaries: general            }
}
\maketitle{}           

%

\section{Introduction}

The limitations imposed by the atmospheric seeing is a serious problem
for ground based observations. Speckle interferometry, which allows one to
circumvent  blurring by the Earth's atmosphere, has been known for three 
decades (Labeyrie 1970) 
and  is mainly applied to the research of close binary and
multiple systems (see the large series of papers by McAlister and 
collaborators), to the measurements of stellar diameters and to the study of the
strucure of circumstellar envelopes at different wavelengths; it has been also
used to evaluate  sizes and shapes of the minor objects of the solar system.
Unfortunately, this technique has not been  widely  applied so far
since  its major limitation lies
in the relatively small dynamic range  allowed for
the object magnitude.
However, 
speckle interferometry, under certain observing conditions,
can still be used
to retrieve the difference in magnitude between objects 
which are quite close
in terms of relative brightness.

In spectral analysis, the flux from a composite object, when interpreted as 
due to a single source, 
will most certainly cause confusion and may originate 
elaborate, but unrealistic, theories. 
Such a confusing situation is  evident in the class of the
\lambo stars, Population I, early-A, recently extended up to early-F 
type stars characterized by metal 
lines much weaker than expected for their spectral type.
The wide range of the derived metal underabundances and the variety of 
explanations of  the \lambo phenomenon are 
found  in the large number of 
recent papers on the identification and interpretation of these stars.  
Faraggiana \& Bonifacio (1999) raised the question 
that undetected duplicity is a possible explanation of the peculiar Balmer
profiles (shallow cores and broad wings) and of the apparent metal 
underabundances of several \lambo candidates; in fact,
in a composite spectrum,  the  veiling effect produces shallow lines 
which are 
characteristic of most \lambo stars (see Corbally 1987).


\section{Observations}

The speckle camera mounted on the Adaptive Optics module (AdOpt@TNG) of 
the 3.5m
Telescopio Nazionale Galileo (TNG) is expected to 
reach  the diffraction limit
($0\farcs{043}$ at 600nm)  and 
is an ideal tool for 
separating narrow binary systems with magnitude differences between their 
components of less than 3 magnitudes, as is expected in
the case of binarity of a \lambo candidate.

The imager is an ICCD Proxitronic camera with a quantum efficiency 
optimized for the blue
part of the visible spectrum ($\approx20$\% at 500nm).
The central part of the TV signal is digitized in a 
$128\times128$ pixel array (8 bits/px) at the standard frame rate of 25 Hz, 
while the single
frame exposure ranges from 2 to 40ms. An optical relay provides a scale of $\approx$
0\farcs{030}/px giving a field of view of $\approx$3\farcs{9}. No atmospheric dispersion
correction is applied.

The speckle camera computes in real--time the power spectrum of each frame and sums directly
the whole set of power spectra obtained during the run. The data are then off--line 
corrected for the instrumental biases such as the background and detector inhomogeneities.
The filter set includes some general purpose (Str\"omgren bands) and some
narrow bandpass ones (e.g. H$_{\alpha}$ or TiO and ZrO  absorption bands).

A detailed description of the real--time speckle facility can be found in
Marchetti 
et al. (1997) and Mallucci (1998), while the real--time data acquisition
is fully described in Baruffolo, Ragazzoni \& Farinato (1998).

A calibration run of the speckle camera of the TNG 
has been used to observe a sample of stars classified as
\lambo from spectroscopic observations; this sample has been extracted 
from the list published by Faraggiana \& Bonifacio (1999) and it is shown in
Table \ref{tab:list}.
We obtained speckle observations of these stars 
on the nights of December 20th and 21st, 1999 and on September 28th, 2000.
The nights were plagued by poor seeing and as a consequence
the signal to noise ratio (SNR) 
for most  observations was not high enough to provide
stringent lower limits for separation and $\Delta m$.
We report here the positive results obtained
for two stars, HD 38545 and HD 290492,
for which  we measured separation, $\Delta m$ and 
position angle, and we give the upper limits we could attain for some other
\lambo candidates. 

The filters 
chosen for the observation were tuned to match both the characteristics
of the objects and the seeing conditions experienced during the two nights.
We decided to use the intermediate band filters $b$ and $y$ of the 
Str\"omgren
system and 
a narrow band H$_{\alpha}$ filter
for the  very bright star HD 38545.

The 
exposure 
time of each speckle frame was 20 ms for all stars
observed, including
those used for the field of view calibration, 
and runs of 3000 frames were performed,
each with a total integration time of 60 seconds per run.
Depending on 
seeing conditions and on the brightness of the target star,
up to 10 runs per object 
were performed.

For each 
object, we selected a reference single star in order to acquire the Speckle
Transfer Function (STF) 
needed to deconvolve the atmospheric disturbance from the power spectrum
of the object. 
Since the behaviour of the seeing is variable with a time scale 
that  may be of the 
order of minutes, and also depends on the zenith distance,
we selected a suitable STF star 
of 
comparable magnitude
within few degrees of each target,
and we switched between them many times, thus allowing  the
best possible homogeneity in terms of temporal seeing variations.

We also selected two double stars having well--known
orbital parameters 
for  determining the detector's scale and 
orientation, namely ADS784 AB and ADS6650 AB,
for which the orbital parameters are taken from 
Cole et al. (1992) and S\"oderhjelm (1999) respectively and
which were observed
with all the three filters mentioned above. 

\begin{center}
\begin{figure}
\psfig{figure=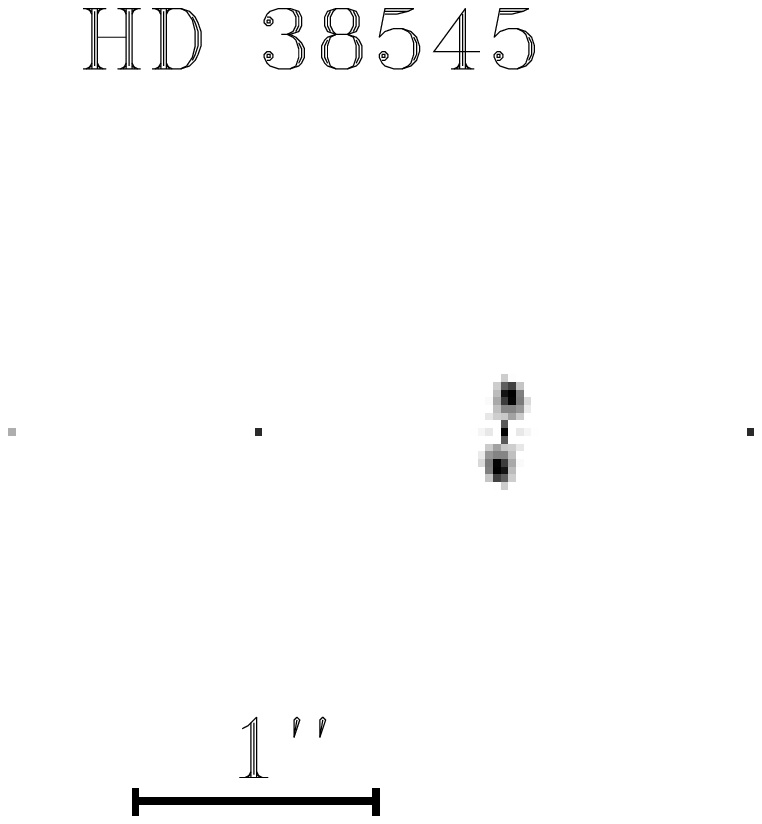,width=8.8truecm}
\caption{Autocorrelation function of HD 38545}
\label{fig:hd38545}
\end{figure}
\end{center}

\section{Data Reduction}

The speckle facility, 
after the end of each run, provides the accumulated power spectrum
of the collected speckle frames. 
The power spectrum is divided by the STF obtained
from observations of a nearby star, 
canceling out in this way the contribution of the atmospheric turbulence
affecting the observation. This image
pre--processing  is
also needed both for 
removing some features caused by the possible repetitive noise induced 
on the camera signal 
and 
to eliminate a typical cross--shaped disturbance occurring when the 
speckle image of the object is 
not entirely contained in the 
camera field of view.
The power spectrum is then 
inverted via Fast Fourier Transform (FFT), and the autocorrelation
function (ACF) of the brightness distribution of the 
astronomical target is obtained.
In the case of binary stars, the 
ACF shows the characteristic behaviour of a central peak
with two opposite and 
symmetric secondary peaks (see Fig.\ref{fig:hd38545}).
The distance between the central peak
and one of the secondary ones 
is the separation between the two components while the position angle
is given by the 
orientation of the secondary peak with  $180^{\circ}$ uncertainty.
The center of the secondary peaks is retrieved by fitting a paraboloid 
with a sub--pixel precision.

The magnitude difference is estimated by comparing the intensities of the
secondary and the central peaks.
The energy contained in the two secondary peaks is computed by integrating
the ACF signal  delimited by the paraboloidal fitting, and the same procedure
is applied to the central peak with its proper paraboloidal fitting.
The central pixel of the ACF is affected by a large amount of spurious signal
given by the correlation of the noise and the background. In the integration process, its value
has been substituted with that estimated by the fitted paraboloid at the same position.
Finally, the comparison between the energies of the secondary peaks and 
that of the central one gives the magnitude difference.
The relative errors are computed using the errors of the paraboloidal fitting.

The unfavourable  weather  conditions 
(poor  seeing and strong wind)
during 
the observations seriously affected the 
instrument performance. 
Even if the speckle interferometry is not as sensible to the seeing as other
high angular resolution 
techniques (i.e. Adaptive Optics), the low SNR achieved 
surely compromised both the possibilty 
to detect very close binary systems (separation
$<0\farcs{1}$), and the accuracy of the magnitude difference measurements.
However, the obtained data are quite 
encouraging and demonstrate that also when seeing conditions
are not 
the most favourable for 
high angular resolution observations, it is still possible to 
attain significant results.

\begin{table}
\caption{Speckle results for HD 38545}
\label{tab:hd38545}
\begin{center}
\begin{tabular}{lccc}
\hline
Filter & Separation & Position angle & $\Delta m$ \\
\hline

$\rm H_\alpha$ &  0\farcs{136} $\pm$ 0\farcs{005} 
& $ 192\fdg{7}  \pm 1\fdg{9} $ & 
$  0.61\pm 0.20 $\\
$y$  & 0\farcs{145} $\pm$ 0\farcs{005} &
$ 191\fdg{4}  \pm 1\fdg{9}   $ &
$ 0.63\pm 0.20  $\\
$b$ & 0\farcs{145} $\pm$ 0\farcs{005} & 
$  189\fdg{5}  \pm 1\fdg{9} $ & 
$0.57 \pm 0.20 $ \\
\hline
\end{tabular}
\end{center}
\end{table}

\section{Results for HD 38545}

This star (=HR 1989 =131 Tau)
was classified as \lambo by Gray \& Garrison (1987)
and, since then, it has been
accepted as belonging to this class by all authors, except for 
Abt \& Morrell (1995) who classified it as a shell star.
   
The observed spectrum mimics
quite well that of a single star, as shown by the abundance analysis by 
St\"urenburg (1993) and by the line profile discussion by Bohlender \& Walker
(1994), who confirm the atmospheric parameters, \teff~
and log g derived by the former author.
The star's shell lines are discussed by Bohlender \& Walker 
(1994), Andrillat et al. (1995), Grady et al. (1996),
Hauck et al. (1998), Holweger et al. (1999), but none of these authors could
find any spectroscopic signature suggesting
that the star is not a single object.

Since this star is quite bright ($V=5.725$)
it was observed with 3 filters: H${_\alpha}$,
$y$ an $b$. Our results  are given
in Table \ref{tab:hd38545} and
the autocorrelation function is shown in Fig.\ref{fig:hd38545}.
There is good agreement between the present results and those  obtained by the 
Hipparcos experiment (separation 0\farcs{155}, $\Delta H_p = 0.64$).

\begin{center}
\begin{figure}
\psfig{figure=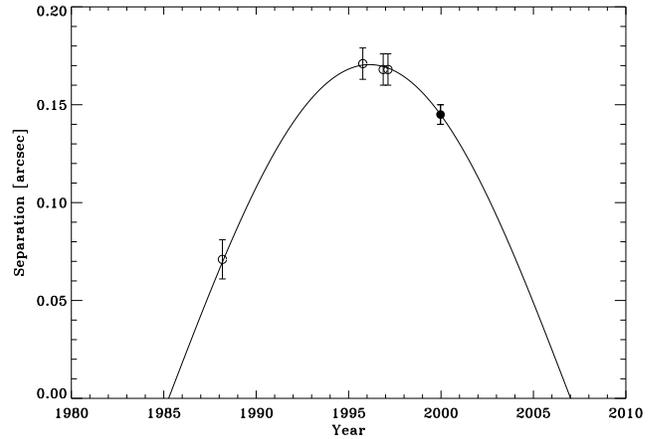,width=8.8truecm}
\caption{Interferometric measurements of HD38545 fitted with a sinusoid: the empty circles are 
the CHARA data and the filled circle is our $y$ measurement. The fitting is made under
the assumption $i=90^{\circ}$ and $e=0$}
\label{fig:orbit}
\end{figure}
\end{center}

The duplicity of HD 38545 was discovered using speckle interferometry 
by McAlister et al. (1993)  and measured again
by  Hartkopf et al. (2000). Of the 4 above measurements of the CHARA group,
the separations in three cases are quite close together ($\approx$0\farcs{170}
in 1995.7686, 1996.8717 and 1997.1311), while the separation  measured
by McAlister et al. (1993) in 1988.1729
is 0\farcs{071}, i.e. over a factor of two smaller than the others.
The position angles measured by the CHARA group are all close
to $190^{\circ}$ and slightly decreasing, as in our case,
and this fact suggests that the orbit is seen nearly edge--on.
Although the amount of data is too small
to retrieve the orbital parameter of this binary, a rough estimation of the orbital
period can be made in the approximate assumption that the inclination is
$i=90^{\circ}$ and the orbit is circular ($e=0$). We fitted the separations vs. the epoch of 
the interferometric observations with a sinusoid and we found a good agreement for $P=43.5$y and
$a=$0\farcs{171} (see Fig.\ref{fig:orbit}). Using the 
Hipparcos parallax from Table \ref{tab:list}
we give an estimation for the total mass of the system 
$M=5.7\pm2.3\;M_{\odot}$.

According to the spectral analysis by St\"urenburg (1993), the two
components are expected to have similar masses of about 2.5 $M_{\odot}$.
In fact, the absolute magnitude of this object corresponds to that
of a star lying more than one magnitude above the ZAMS if
the duplicity is not taken into account, and therefore its position 
on the HR diagram given by Paunzen (1997), Paunzen et al. (1998)
(who also computed a wrong value of M$_{V}$) and, on the 
colour-magnitude diagram, by Bohlender et al. (1999) is misleading.

\section{Results for  HD 290492}

The characteristics of this binary system reported in the Washington 
Double Star (WDS) catalogue
are  $\Delta m=1.4$ and d=0\farcs{6}, while
medium-resolution spectroscopic observations 
allowed Paunzen \& Gray (1997) to resolve the system; in fact 
they claim to have measured $\Delta m=0.9$ and 
a separation of $2 ''$. 
The data of the WDS catalogue are based on 3 visual observations made by 
R.A. Rossiter at the Lamont-Hussey Observatory of
Bloemfontein, South Africa, with a 27 1/2-inch refractor 
especially constructed by Zeiss for double star observation.
The data from Rossiter (1955) are 
collected in Table \ref{tab:rossiter}, and demonstrate the 
remarkable accuracy which may be obtained visually
by an experienced observer
 with an appropriate instrument. 

The star is not present in the Hipparcos catalogue, but is found in 
the Tycho catalogue. The transit data (37 accepted
transits for the photometry) have been searched
for binarity but none was detected. There is also
no variability flag, although the scatter in the $V_T$ magnitude
is 0.234.

We observed the binary only with a Str\"omgren $b$ filter and our results 
are summarized in Table \ref{tab:hd290492}.
There is  good agreement with the measurements of Rossiter,
but not with the estimate of Paunzen \& Gray (1997).

\begin{table}
\caption{HD 290492 data from the literature}
\label{tab:rossiter}
\begin{center}
\begin{tabular}{lccc}
\hline
Date & Separation & Position angle & $m_1$--$m_2$ \\
\hline
$1943.026$ & 0\farcs{64} & 65\fdg{6} & 9.8--11.2\\
$1943.205$ & 0\farcs{60} & 69\fdg{4} & 9.8--11.1\\
$1950.174$ & 0\farcs{64} & 67\fdg{8} & 9.7--11.2\\
\hline
\end{tabular}
\end{center}
\end{table}

\begin{table}
\caption{Speckle results for HD 290492}
\label{tab:hd290492}
\begin{center}
\begin{tabular}{lccc}
\hline
Filter & Separation & Position angle & $\Delta m$ \\
\hline
$b$ & 0\farcs{739} $\pm$ 0\farcs{005} & 
$  63\fdg{9}  \pm 1\fdg{9} $ & 
$0.63 \pm 0.20 $ \\
\hline
\end{tabular}
\end{center}
\end{table}

\section{Stars observed with negative result}

Other \lambo candidates have been observed in poor weather conditions.
For all the observed stars
we  list in Table \ref{tab:list} the parallax and its error given by the 
Hipparcos
catalogue and, in the last column, the upper limit
on the possible separation derived from considerations on
seeing and SNR; the error is evaluated to be about
$\pm$ 10 mas. For the two stars for which
the separation has been measured, this upper limit is smaller
than the measured separation. Taking into account the degradation of the
fringe contrast in the object power spectra due to the bad seeing,
we were not able to separate
stars closer than $\approx$0\farcs{10} also considering the brightest
objects and/or a small magnitude difference between the components.
They deserve further observations for more stringent separation
values.  

   \begin{table}
      \caption[]{Upper limits on the separation achieved for each of
the program stars; the separation of the companion of the stars marked with
${\star}$ is given in the previous sections}
         \label{tab:list}
        \begin{center}
        \begin{tabular}{crcccc}
            \hline
            \noalign{\smallskip}
 HD  & 
V & $\pi$  & $\sigma(\pi)$  & upper limit \\
 & & & (mas) & (mas)  \\
            \noalign{\smallskip}
            \hline
            \noalign{\smallskip}

  3\phantom{$^{\star}$A}      & 6.70 & 6.66  & 0.75 &  155 \\
  11503\phantom{$^{\star}$A}  & 4.64 & 15.96 & 0.85 &  124 \\
  23392\phantom{$^{\star}$A}  & 8.7  & 3.25  & 1.08 &  310 \\
  38545$^{\star}$\phantom{A}  & 5.72 & 7.72  & 0.93 &  124 \\
  39421\phantom{$^{\star}$A}  & 5.97 & 8.60  & 0.92 &  124 \\
  64491\phantom{$^{\star}$A}  & 6.23 & 16.55 & 0.92 &  124 \\
  74873\phantom{$^{\star}$A}  & 5.87 & 16.38 & 1.16 &  124 \\
  84123\phantom{$^{\star}$A}  & 6.81 & 9.09  & 0.90 &  155 \\
  84948\phantom{$^{\star}$A}  & 8.1  & 4.97  & 1.14 &  284 \\
  90821\phantom{$^{\star}$A}  & 9.2  & ---   & ---  &  310 \\
  91130A\phantom{$^{\star}$}  & 5.93 & 13.33 & 0.76 &  124 \\
  98772\phantom{$^{\star}$A}  & 5.98 & 11.58 & 0.56 &  124 \\
  105058\phantom{$^{\star}$A} & 8.91 & 5.32  & 1.04 &  310 \\
  153808\phantom{$^{\star}$A} & 3.92 & 20.04 & 0.65 &   93 \\
  192640\phantom{$^{\star}$A} & 4.97 & 24.37 & 0.55 &  124 \\
  204041\phantom{$^{\star}$A} & 6.46 & 11.46 & 0.99 &  155 \\
  290492$^{\star}$\phantom{A} & 9.27 & --    & --   &  310 \\
         \noalign{\smallskip}
         \hline
         \end{tabular}
         \end{center}
  \end{table}

We add here only a few comments on the duplicity of HD 153808, 
the star which has the lowest upper limit on the possible separation
of a companion.
Controversial visual 
binary detections are reported in the literature for this star. 
Its duplicity is measured by Isobe et al. (1990, 1992) from
speckle observations, which are not confirmed by other authors
(Miura et al. 1992, 1995; McAlister et al. 1993; Kuwamura et al. 1993).
This star has been observed by Hipparcos, but no sign of duplicity
has been detected and no mention of its duplicity is given in the
Hipparcos Input Catalogue (Turon et al. 1993).

It is discussed as spectroscopic binary by
Petrie (1939) who classified the two components as A0 and A2, 
computed a magnitude difference of 1.5 and showed, in Fig. 5 of his paper, the
line profiles of three lines at different phases.
Batten et al.'s (1989) catalogue gives the orbital elements 
($ a \sin i$ being $3.91\times 10^6$ and $6.2\times 10^6$ km
for the two components) and Hipparcos measured the parallax 
$\pi=20.04\pm 0.65$ mas. 
According to these 
data the angular separation of the two components of the spectroscopic
binary system should be not higher than 0.13 mas 
and so these stars cannot be identified 
with those detected by the speckle observations. This low expected value of
the angular separation  explains the lack of duplicity detection
by the Hipparcos experiment as well as by our TNG observations. 
This demonstrates   that 
when the the separation is too small to be detected by direct imaging
and the  spectral lines are too broad
to separate the components by spectroscopic observations,
it is impossible to establish the binary nature of a system.  

\section{Discussion }

The characteristics of the \lambo stars are still not yet explained, in spite
of the numerous efforts made, especially in the last two decades. The 
inhomogeneous properties of the members of this class represent the most
intriguing aspect of the problem and the large area these stars occupy
on the HR diagram represents a serious problem for the determination of
their evolutionary stage.
We recall that no systematic search for binaries has been made for these objects
and we consider this point as the first to be clarified before any study
can be initiated (see Faraggiana \& Bonifacio, 1999).

Our search for binaries with the TNG speckle camera has been severely
limited by poor weather conditions, but it allowed us to confirm that
two \lambo stars HD 38545 and HD 290492 must be removed from this class
of objects.

Before making any detailed analysis of peculiar objects and  elaborating
theories on their characteristics, a rigorous selection of true single
objects is required.
This may prove to be impossible from spectroscopic data alone.

For example, the duplicity of HD 38545 
is now well established, thanks to speckle and astrometric observations,
while it had never been suspected from the
several analyses of its spectrum. 

The case of HD 153808 represents an opposite example:
high quality spectra
revealed, over 60  years ago, that the star is a binary, while the present
speckle observations and the Hipparcos experiment
did not succeed in detecting its duplicity. 
The dubious visual duplicity found by previous speckle observations may
suggest the presence of a third body, which,
however, cannot be responsible for the
SB2 system. 

These two objects clearly show that
a single best method to detect binaries does not exist; 
this is confirmed by the fact that the positive duplicity detection 
has been obtained for two stars which are not 
the brightest, nor the nearest objects (see V and $\pi$ values in Table 4)
and not even those observed under the best conditions.
We cannot guess which observing approach, direct imaging or spectroscopy,
is more suitable for duplicity detection; 
only coordinated
efforts using different observational techniques will be efficient in revealing
new binaries which produce a composite spectrum.

\section { Conclusions }

We performed a search of duplicity among \lambo candidates
using the speckle camera of the Galileo telescope.
We have been able to confirm the  separation and
$\Delta m$ for  two of the
program stars; for the others we were able to place stringent
upper limits on the separation of a possible companion.
The use of this instrumentation is promising mainly
because it allows the determination of both separation
and $\Delta m$, which is not always possible
by  the speckle approach. 
Due to the poor weather conditions we were not able
to assess if the theoretical diffraction
limit may be actually achieved nor could we 
establish the limiting magnitude and
maximum $\Delta m$ for successful binary detection.

Although we have shown that the speckle camera can work
even under bad weather conditions, the observations would
greatly benefit from a good seeing. 
In such conditions, the speckle camera should allow to
reach an angular resolution which is almost an order of magnitude
better than that obtained by classical ground based instruments
and comparable with that of space instrumentation. 

\begin{acknowledgements}
We thank the Referee, Dr. Y.Y. Balega, for the useful suggestions and for 
pointing out a serious error in the first version of the manuscript.
We also thank Dr. R. Ragazzoni for the useful discussions of the results
and Dr. A. Ghedina for the invaluable support during the observations.

\end{acknowledgements}

\end{document}